\documentclass[pre,12pt,onecolumn,showpacs]{revtex4}

\usepackage{epsfig}
\usepackage{amssymb,amsmath,stmaryrd,tabularx}
\usepackage{wrapfig}
\usepackage{subfigure}
\usepackage{bm}
\usepackage{graphicx}
\newcommand{\bl}{\left \llbracket}
\newcommand{\br}{\right \rrbracket}
\begin{document}
\title{The thermodynamics and roughening of solid-solid
interfaces}

\author{Luiza Angheluta, Espen Jettestuen, and Joachim Mathiesen}
\affiliation{Physics of Geological Processes, University of Oslo, Oslo, Norway }
\date{\today}
%
%

\begin{abstract}
The dynamics of sharp interfaces separating two non-hydrostatically
stressed solids is analyzed using the idea that the rate of mass
transport across the interface is proportional to the thermodynamic
potential difference across the interface. The solids are allowed to
exchange mass by transforming one solid into the other,
thermodynamic relations for the transformation of a mass element are
derived and a linear stability analysis of the interface is carried
out.  The stability is shown to depend on the order of the phase
transition occurring at the interface. Numerical simulations are
performed in the non-linear regime to investigate the evolution and
roughening of the interface. It is shown that even small contrasts
in the referential densities of the solids may lead to the formation
of finger like structures aligned with the principal direction of
the far field stress.
\end{abstract}

%
%

\maketitle

%
%

\section{Introduction}
The formation of complex patterns in stressed multiphase systems is
a well known phenomenon. The important studies of Asaro and
Tiller~\cite{Asaro72} and Grinfeld~\cite{Grinfeld86} brought
attention to the morphological instability of stressed surfaces in
contact with their melts or solutions. In the absence of surface
tension, small perturbations of the surface increase in amplitude
due to material diffusing along the surface from surface valleys,
where the stress and chemical potential is high, to surrounding
peaks where the stress and chemical potential is low. Important
examples of instabilities at fluid-solid interfaces include defect
nucleation and island growth in thin films ~\cite{Srolovitz88,
Gao99}, solidification~\cite{Mullins57} and the formation of
dendrites and growth of fractal clusters by
aggregation~\cite{Meakin98}. The surface energy increases the
chemical potential at regions of high curvature (convex with respect
to the solution or melt, at the peaks) and reduces the chemical
potential at region of low curvature (at the valleys) and this
introduces a characteristic scale below which the interface is
stabilized.

In systems where the fluid phase is replaced by another solid phase,
i.e. solid-solid systems, the interface constraints alter the local
equilibrium conditions. Here we study a general model for a
propagating interface between non-hydrostatically stressed solids.
The interface propagates by mass transformation from one phase into
the other. The phase transformation is assumed to be local, i.e. the
distance over which the solid is transported via surface diffusion
or solvent mediated diffusion is negligible compared to other
relevant scales of the system. Although the derivations apply to a
diffuse interface, we shall here treat only coherent interfaces,
where there is no nucleation of new phases or formation of gaps
between the two solids ~\cite{Robin74,Larche78}, in the sharp
interface limit. For example, in rocks such processes appear at the
grain scale in "dry recrystallization"~\cite{Kamb58, Fletcher73}.
Common examples of coherent interfaces that migrate under the
influence of stress include 
the surfaces of
coherent precipitates (stressed inclusion embedded in a crystal
matrix) \cite{Robin74} and interfaces associated with isochemical
transformations. Most studies of solid-solid phase transformations
have been limited to the calculation of chemical potentials in
equilibrium and have provided little insight into the kinetics. Here
we investigate the out of equilibrium dynamics of mass exchange
between two distinct solid phases separated by a sharp interface. We
expand on the recent work presented in \cite{PRL08} where we studied
the phase transformation kinetics controlled by the Helmholtz free
energy. It was shown that a morphological instability is triggered
by a finite jump in the free energy density across the interface,
and in the non-linear regime this leads to the formation of finger
like structures aligned with the principal direction of the applied
stress.

\begin{figure}
\epsfig{file=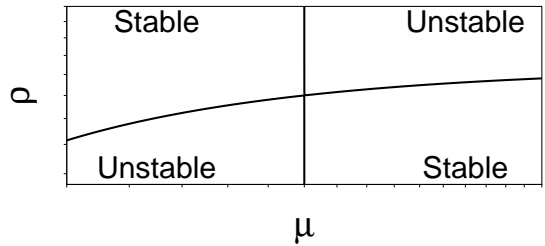,width=.48\textwidth}
\caption{Sketch of a stability diagram for the growth rate of a
sharp interface separating two solid materials. The axes show
relative values of the shear modulus and density of the phases. As
it will be shown in Sec. III, the symmetry of the diagram is broken
by the values of the Poisson's ratios.}\label{stabsketch}
\end{figure}

In the majority of solid-solid phase transformation processes, the
propagation of the interface is accompanied by a change in density.
For this reason the density is an important order parameter that
quantitatively characterizes the difference between the two phases.
We consider two types of phase transitions underlying the kinetics,
first order and second order, which result in fundamentally
different behaviors at the phase boundary. A first order phase
transition occurs when the two phases have different referential
densities and it typically results in morphological instability
along the boundary whereas a second order phase transition may
either stabilize or destabilize the interface depending on Poisson's
ratios of the two phases. A simple sketch of the stability diagram
is outlined in Fig. \ref{stabsketch} for relative values of density
and shear modulus of the two phases.

The article consists of five sections. In Sec. II we derive a
general equation for the kinetics for mass exchange at a solid-solid
phase boundary separating two linear elastic solids. We utilize the
derived equations on a simple one dimensional example and offer a
short discussion of the order of the phase transition underlying the
kinetics. We proceed in Sec. III with a linear stability analysis of
the full two-dimensional problem. In two dimensions, the phase
transformation kinetics gives rise to the development of complex
patterns along the phase boundary. While we solve the problem
analytically for small perturbations of a flat interface, things
become more complicated in the non-linear regime, and we resort to
numerical simulations based on the combination of a Galerkin finite
element discretization with a level-set method for tracking the
phase boundary. In Sec. IV, numerical results are presented together
with discussions. Finally in Sec. V we offer concluding remarks.

\section{General phase transformation kinetics}

\begin{figure}
\epsfig{file=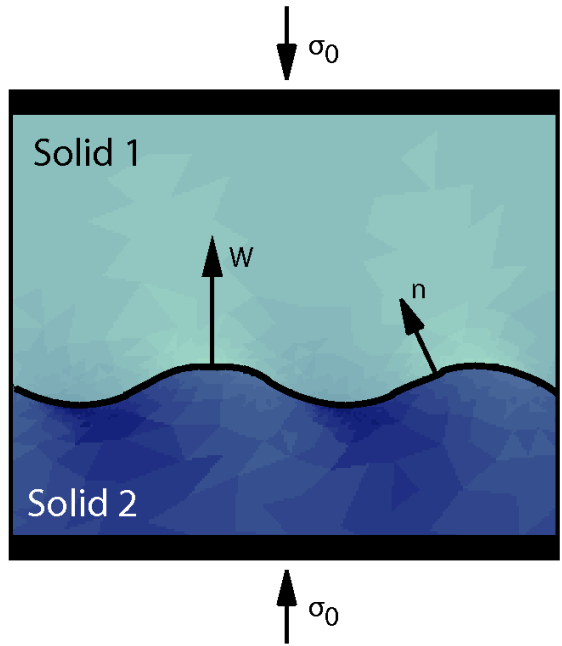,width=.45\textwidth}
\caption{(color online) Two solids separated by a sharp interface. A
compressional force is applied at the margins in the vertical
direction}\label{2dsetup}
\end{figure}

Although the equations that we derive for the exchange of a mass
element between two solid phases in a non-hydrostatically stressed
system apply to more general settings, we limit ourselves to the
study of two solids separated by a single sharp interface. The
solids are stressed by an external uniaxial load as illustrated in
Fig. \ref{2dsetup}. In the referential configuration, a solid phase is
assumed to have a homogenous mass density, $\rho^0$, defined per
unit undeformed volume occupied by that phase. After the deformation, the densities are
functions of space $x$ and time $t$, i.e. $\rho_1(x,t)$ and
$\rho_2(x,t)$. The average density of the two-phase system is denoted by $\rho(x,t)$. Finally, the mass fraction for
phase $1$ is denoted by $c$. In this notation, the mass fraction of
phase $2$ becomes $1-c$.

For non-vanishing densities, the mass-averaged velocity is defined as
\begin{eqnarray}\label{vbar}
\bar v = cv_1+(1-c)v_2.
\end{eqnarray}
Throughout the text, the mass average of any quantity is indicated
by a bar. Similarly, the average specific free energy density is
given by
\begin{eqnarray}\label{fbar}
\bar f = cf_1+(1-c)f_2.
\end{eqnarray}
The total specific volume is related to the real densities in the
deformed state, $\rho_1(x,t)$ and $\rho_2(x,t)$ by
\begin{equation}\label{rho-1}
\rho^{-1} = c\rho_1^{-1}+(1-c)\rho_2^{-1}.
\end{equation}

The interface separating the two phases is tracked by the zero level
of a scalar field $\phi(x,t)$ passively advected according to the
equation
\begin{equation}\label{eq:ls2}
\frac{\partial\phi}{\partial t} + W\left|\nabla\phi\right| = 0,
\end{equation}
where $W$ is the normal velocity of the surface. It follows that the
interface is  given by the zero level set
\begin{equation}
\Gamma = \left\{x|\phi(x,t) = 0\textrm{, for all }t\right\}.
\end{equation}
The scalar field is constructed such that phase 1 occupies the
domain in which $\phi(x,t)>0$ and phase 2 occupies the domain in
which $\phi(x,t)<0$, see Fig. \ref{2dsetup}. In this notation, the
mass fraction may be expressed as the characteristic function of the
scalar field,
\begin{eqnarray}
c(x,t)=H\left(\phi(x,t)\right) =
\left\{\begin{array}{c}1\textrm{, if }\phi(x,t)>0\\\frac{1}{2}\textrm{, if }
\phi(x,t) = 0\\0\textrm{, otherwise.}\end{array} \right.
\end{eqnarray}
In the subsequent analysis, we make use of the following relations (see e.g. \cite{Trusk98})
\begin{gather}
\nabla_i c = n_i\delta_\Gamma, \quad \partial_t c = -W\delta_\Gamma,
\end{gather}
where $n_i = {\nabla_i\phi}/{|\nabla\phi|}$ is the normal unit
vector of the interface, $W=-{\partial_t\phi}/{|\nabla\phi|}$ is the
normal velocity and $\delta_\Gamma = |\nabla\phi|\delta(\phi)$ is
the surface delta function.

Taking the gradient of the averaged velocity from Eq. (\ref{vbar})
and using the above identities, the following relation is obtained
\begin{eqnarray}
\nabla_i \bar v_j &=& \frac{\partial\bar v_j}{\partial c}\nabla_i c+c\nabla_iv_{1,j}+(1-c)\nabla_iv_{2,j}\nonumber\\
&=& \frac{\partial\bar v_j}{\partial c}n_i\delta_\Gamma+\overline{\nabla_iv_j}.\label{gradv}
\end{eqnarray}

\subsection{Kinetics of the phase transformation}
The system must satisfy fundamental conservation principles for the
mass, momentum, energy and entropy. Let us denote the material time
derivative with respect to the mass-averaged velocity by a dot, i.e.
$\dot{\Theta} = \partial_t\Theta+\bar v_i\nabla_i\Theta$. Then, the
local mass conservation can be written in the form
\begin{equation}\label{rhodot}
\dot\rho = -\rho\nabla_i\bar v_i.
\end{equation}
and the local momentum balance can be written in the form
\begin{equation}\label{vdot}
\rho\dot{\overline{v}_i} = \nabla_j\sigma_{ij},
\end{equation}
where $\sigma_{ij}$ is the stress tensor.

The mass fraction of phase 1 satisfies the advection-reaction equation given by
\begin{equation}\label{cdot}
\rho \dot c = Q\delta_\Gamma,
\end{equation}
where the mass exchange rate $Q$ is confined to the interface by the
delta-function (in the sharp interface limit). Mass transport by
diffusion is negligible in the reaction dominated regime. This is a
valid approximation when the characteristic length $\ell =
\cal{D}/W$, where $\cal D$ is the diffusion coefficient and $W$ is
the velocity of the interface, is small compared with other relevant
microscopic length scales. That is material diffusion occurs on a
time scale much longer than any other relevant time scale in the
system or equivalently the characteristic length scale formed from
the diffusion constant and solidification or precipitation rate is
small compared to other relevant microscopic scales.

In the linear kinetics, the mass exchange rate is now derived from
the requirement that the entropy production has a positive quadratic
form. We start by expressing the conservation of specific energy
density $e$ in the form
\begin{equation}\label{edot}
\rho\dot{\bar e} = \sigma_{ij}\nabla_i\bar v_j,
\end{equation}
where $\bar v^2 = c v_1^2+(1-c) v_2^2$ since the cross term vanishes
in the limit of a sharp interface.

At equilibrium
\begin{equation}\label{bare}
\bar e = \bar f+T\bar s,
\end{equation}
where the free energy is assumed to be a function of the local
strain and the composition, i.e. $\bar f = \bar
f(\bar\epsilon_{ij},c)$. By inserting the energy conservation
equation, Eq. (\ref{edot}), into the time derivative of this
equation, under constant temperature conditions, the expression
\begin{equation}
\rho T\dot{\bar s} = \sigma_{ij}\nabla_i \bar v_j-\rho\frac{\partial\bar f}{\partial\bar\epsilon_{ij}}\dot{\overline{\epsilon_{ij}}}-\rho\frac{\partial \bar f}{\partial c}\dot c,
\end{equation}
is obtained. The phase transformation is assumed to be slow and
isothermal. From Eqs. (\ref{fbar}) and (\ref{gradv}) it follows that
\begin{equation}
\rho T\dot {\bar s} = \sigma_{nj}\frac{\partial \bar v_j}{\partial c }\delta_\Gamma+\sigma_{ij}\overline{\nabla_iv_j}-\rho\frac{\partial\bar f}{\partial\bar\epsilon_{ij}}\dot{\overline{\epsilon_{ij}}}-\frac{\partial f}{\partial c}\rho\dot c.
\end{equation}
Given that the strain rate is $\dot\epsilon_{ij}=1/2(\nabla_i
v_j+\nabla_j v_i)$ and using the symmetry of the stress tensor, we
arrive at the expression
\begin{equation}
\rho T\dot {\bar s} = \sigma_{nj}\frac{\partial \bar v_j}{\partial c}\delta_\Gamma+\left(\sigma_{ij}-\rho\frac{\partial\bar f}{\partial\bar\epsilon_{ij}}\right)\dot{\overline{\epsilon_{ij}}}-\frac{\partial f}{\partial c}Q\delta_\Gamma ,\label{let}
\end{equation}
where $\sigma_{nj}=\sigma_{ij}n_i$ is the stress vector at the
interface. From Eqs. (\ref{gradv}) and (\ref{rhodot}) and using an
equation of state of the form $\rho(\bar\epsilon_{ij},c) =
\rho^0(c)(1-\bar\epsilon_{ii})$ it follows that,
\begin{eqnarray*}
\frac{\partial \rho}{\partial c} \dot c +\frac{\partial\rho}{\partial\bar\epsilon_{ij}}\dot{\overline{\epsilon_{ij}}}&=& -\frac{\partial v_n}{\partial c}\rho\delta_\Gamma-\rho\overline{\nabla_i v_i} \Rightarrow \\
\frac{1}{\rho}\frac{\partial\rho}{\partial c}Q\delta_\Gamma -\rho^0\dot{\overline{\epsilon_{ii}}}&=& -\frac{\partial v_n}{\partial c}\rho\delta_\Gamma-\rho\overline{\nabla_i v_i}\Rightarrow\\
\frac{\partial}{\partial c}\left(\frac{1}{\rho}\right) Q = \frac{\partial v_n}{\partial c} &,\quad& \rho^0\dot{\overline{\epsilon_{ii}}}\approx\rho\overline{\nabla_iv_i}.
\end{eqnarray*}
Using Eq. (\ref{rho-1}) for the density, the jump in the material
velocity is related to the reaction rate by
\begin{equation}
\frac{\partial v_n}{\partial c} = Q\frac{\partial}{\partial c}\left(\frac{1}{\rho}\right).
\end{equation}
The direction of the kinetics is constrained by the second law of
thermodynamics which can be expressed in the continuum form as
\begin{equation}\label{sdot}
\rho\dot \bar s +\nabla_iJ^s_i=\Pi_s,
\end{equation}
where $J^s_i$ is the entropy flux density and $\Pi_s\ge 0$ is the
entropy production rate. We consider the case where the entropy flux
is negligible (in the absence of mass and heat fluxes) and therefore
set $J_s = 0$. Combining Eqs. (\ref{let}) and (\ref{sdot}), it can
be seen that the positive entropy production rate leads to the
condition
\begin{equation}
\left(\sigma_{nn}\frac{\partial }{\partial c}\left(\frac{1}{\rho}\right)
-\frac{\partial \bar f}{\partial c}\right)Q\delta_\Gamma+\left(\sigma_{ij}
-\rho\frac{\partial\bar f}{\partial\bar\epsilon_{ij}}\right)\dot{\overline{\epsilon_{ij}}}=
T\Pi_s\ge 0\label{entp}
\end{equation}
on the reaction rate. We now define a constitutive relation that
couples the stress to the strain via the Helmholtz free energy,
\begin{equation}
\sigma_{ij} = \rho\frac{\partial \bar f}{\partial\bar\epsilon_{ij}}.
\end{equation}
From Eq. (\ref{entp}) we observe that the entropy is produced only
at the interface, and in the linear kinetics regime the reaction
rate is proportional to (see e.g. \cite{entprod}),
\begin{equation}\label{Q}
Q\approx K \left(\sigma_{nn}\frac{\partial }{\partial c}\left(\frac{1}{\rho}\right)
-\frac{\partial \bar f}{\partial c}\right),
\end{equation}
where $K>0$ is a system specific constant.

The normal velocity of a sharp interface is obtained by integrating
Eq. (\ref{cdot}) across the interface and taking the singular part
of it,
\begin{eqnarray}
W \approx \bar v_n-\frac{K}{\rho}\bl\sigma_{nn}\frac{1}{\rho}-f\br.\label{nvec}
\end{eqnarray}
Here we introduce the jump in the quantity $a$ from one phase to
another $\bl a\br:=a_1-a_2$, where $a_i$ is the value of $a_i$ in
phase $i$ outside the interface zone as the interface is approached.
The additional interfacial jump conditions of the total mass and
force balance from Eqs. (\ref{rhodot}) and (\ref{vdot}) are given by
\begin{eqnarray}
\llbracket \rho(W-v_n)\rrbracket =0\\
\llbracket \sigma_{ij}n_j\rrbracket =0.
\end{eqnarray}

In general, surface energy $\gamma$ and surface stresses may have an
important effect on the kinetics at the phase boundary with high
curvature $\mathcal K$, therefore the expressions given above are
modified to take this into account. For this purpose we utilize the
Cahn-Hilliard formalism \cite{Cahn58} of a diffuse interface. The
surface energy is obtained by allowing the Helmholtz free energy
density to be a function of the mass fraction gradients, i.e.
\begin{equation}
\rho\bar f(\bar\epsilon_{ij},c,\nabla c) = \rho\bar f_0(\bar\epsilon,c)+\frac{\kappa_1}{2}|\nabla c|^2,
\end{equation}
where $\kappa_1$ is a small parameter related to the infinitesimal
thickness of the interface and $\bar f_0$ is the homogenous free
energy density introduced above. Because the composition gradient is
small everywhere except for a thin zone at the interface, the free
energy can be separated into bulk and surface contributions. If we
now take the limit of vanishing surface thickness and follow the
derivations in the appendix we obtain the general jump condition for
the normal force vector,
\begin{eqnarray}\label{forcejump}
 \llbracket\sigma_{nn}\rrbracket =-2\mathcal K\gamma.
\end{eqnarray}
In the aforementioned expression of the interfacial velocity
Eq.~(\ref{nvec}) the normal stress vector was continuous across the
interface. In the presence of surface tension, the normal velocity is
altered by an additional contribution from the
surface energy,
\begin{equation}
W \approx v_{1,n}+\frac{K}{\rho_1}\left(\bl
  f\br-\langle\sigma_{nn}\rangle\bl\rho^{-1}\br+2\mathcal{K}\gamma\langle
  \rho^{-1}\rangle \right),\label{massex}
\end{equation}
where we have used the interface average defined as $\langle a\rangle =1/2(a_1+a_2)$.

%
%

\subsection{Example: Phase transformation kinetics in a one dimensional system}
\begin{figure}
\epsfig{width=.48\textwidth,file=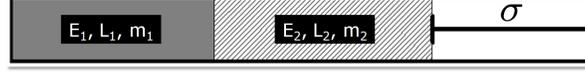}
\caption{One dimensional system undergoing phase transformation}\label{fig1d}
\end{figure}

We start out considering the phase transformation kinetics of a one
dimensional system composed of two linear elastic solids separated
by a single interface. A force $\sigma$ is applied at the boundary
of the system (see Fig. \ref{fig1d}) and each solid phase is
represented by its Young's modulus $E_i$ ($i=1,2$), undeformed
density $\rho_i^0$ and length $L_i^0$. In the deformed state when
the external force is applied the length becomes $L_i = L_i^0 \left
(1+ \sigma /{E_i} \right)$ and the density
$\rho_i=\rho_i^0L_i^0/L_i$. The specific free energy is given by
\begin{equation}
f = \frac{\sigma^2}{2}\left (
\frac{c}{\rho_1(E_1+\sigma)}+\frac{1-c}{\rho_2(E_2+\sigma)}\right).
\end{equation}
In the following, we do not allow new phases to nucleate within the
solids and limit our considerations to the propagation of a single
interface separating the solids. The system is assumed to be
isothermal and no diffusion of mass takes place. The interface moves
as one phase, slowly transforms into the other and an amount
$\rho_1dL_1$, of solid 1 is replaced by an amount $\rho_2dL_2$ of
solid 2, with conservation of the total mass. The phase
transformation is assumed to be irreversible and to occur on time
scales that are much larger than the time it takes for the system to
relax mechanically under the deformational stresses.

In the one dimensional setting the local mass exchange rate is given
by a linear kinetic equation, Eq.~(\ref{Q}), of the form
\begin{equation}\label{reacq}
\dot m_1 = -K\bl\frac{\sigma^2}{2\rho^0E}-\frac{\sigma}{\rho}\br=
K\bl\frac{\sigma^2}{2\rho^0
  E}+\frac{\sigma}{\rho^0}\br,
\end{equation}
with $K>0$. In most cases, the contribution from the jump in the
elastic energy density will be small compared to the contribution
from the work term (because $\sigma/E\ll 1$, within the linear
elasticity regime). The change in the total length will in general
follow the sign of the stress
\begin{eqnarray*}
  \dot L &=& \dot L_1 (1-\frac {\rho_1}{\rho_2})=\dot m_1\bl\frac 1 {\rho} \br\\
&=&K\bl\frac {\sigma^2} {2E\rho^0}+\frac \sigma {\rho^0}\br\bl\frac
1 {\rho^0}+\frac \sigma {E\rho^0}\br.
\end{eqnarray*}
If the densities in the undeformed states are
identical, $\rho_1^0=\rho_2^0$, the change in the total length is given by
\begin{equation}
\dot L=K\frac {\sigma^3} {2\rho^0} \bl\frac 1 { E}\br^2,
\end{equation}
whereas a jump in the referential densities ($\rho^0_1\neq\rho^0_2$)
will result in a work term given by
\begin{equation}
\dot L\approx K\sigma \bl\frac 1 {\rho^0}\br^2.
\end{equation}
Under a compressional load, the dense phase grows at the expense of
the less dense phase (if the two phases have the same Young's
modulus) and the soft phase grows at the expense of the hard phase
(if the two phases have the same density), such that overall the
system responds to the external force by shrinking. The
one-dimensional model cannot predict the morphological stability of
the propagating phase boundary in two dimensions. It turns out that
the work term destabilizes the propagating boundary under a
compressional load.

\subsection{First and second order phase transitions: Equilibrium phase diagrams}
\begin{figure}
\epsfig{width=.48\textwidth,file=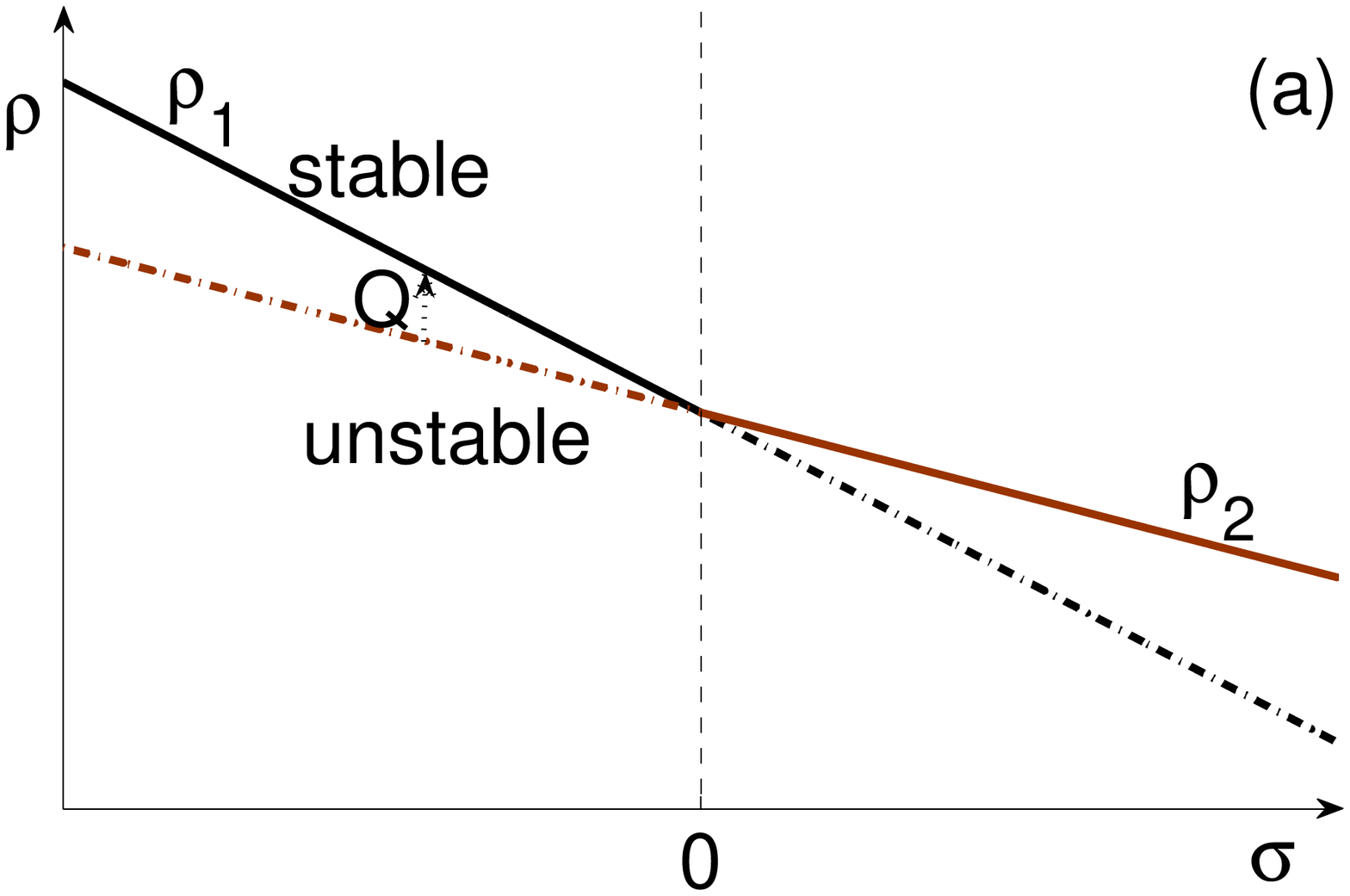}
\epsfig{width=.48\textwidth,file=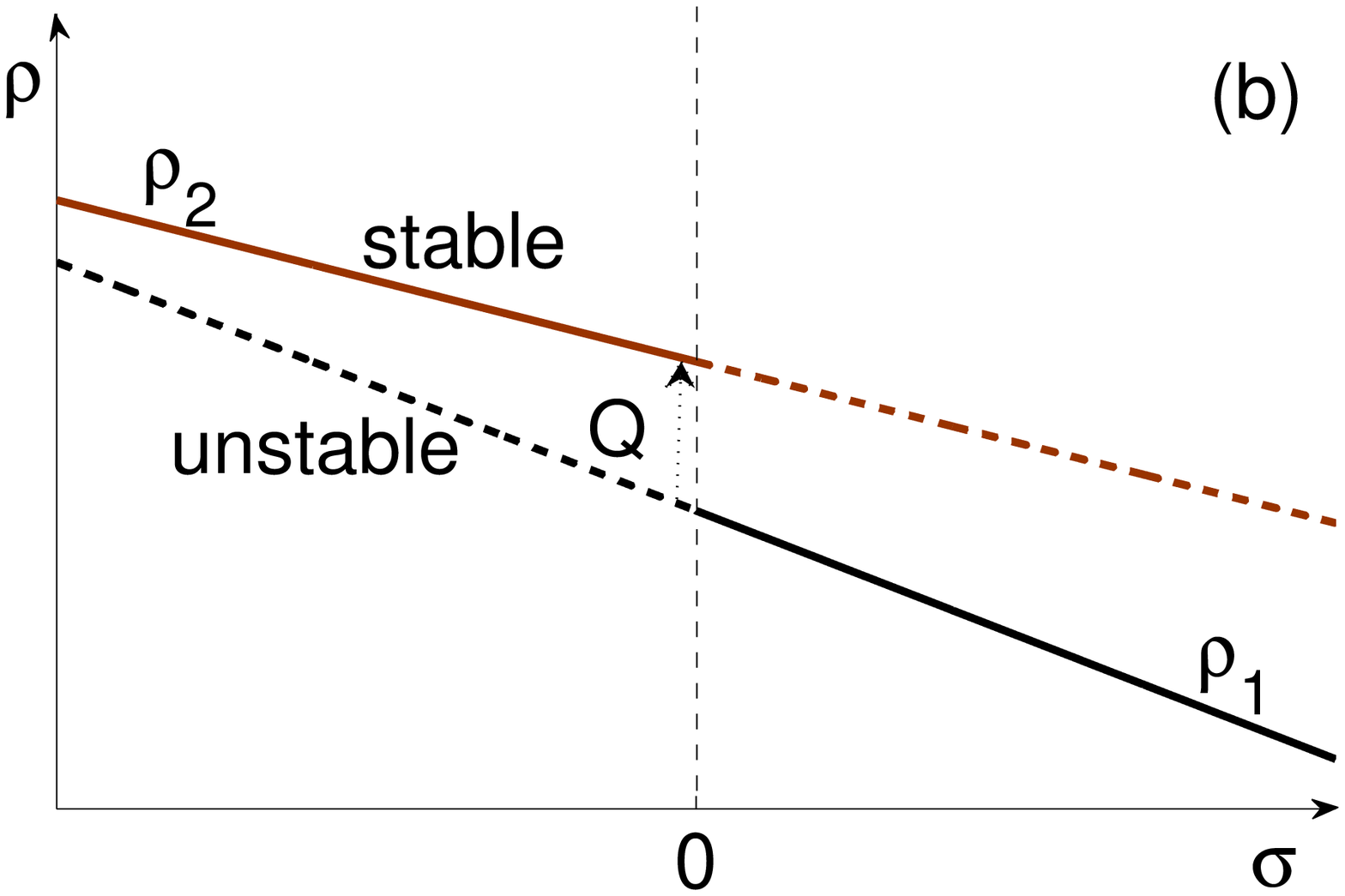}
\caption{(color online) Part (a) illustrates the phase diagram for a
second order phase transition in the $\rho-\sigma$ plane. The
solid-solid kinetics will always be directed from the unstable phase
(dashed line) to the stable phase as illustrated by reaction path
$Q$ marked by the dashed arrow. The slopes of the densities with
respect to stress are Young's modules of the materials. Part (b)
illustrates the equilibrium curves of the first order phase
transition. For the first order phase transition one would in general expect to see hysteresis effects extending the continuous lines (stable regions) beyond the point $\sigma=0$. Here we have shown an idealized case where such effects are disregarded.}\label{figphase}
\end{figure}

In the above derivations, the reaction rate is determined by the
jump in the Gibbs potential across the phase boundary. Whenever the
system is stressed, only one of the two phases will be stable, i.e.
the general two phase system will always evolve to an equilibrium
state consisting of a single phase. In the absence of an external
stress, it is possible for two phases to coexist without any phase
transformation taking place. In the one dimensional example, the
relevant field variable is the stress $\sigma$ applied to the system
and the Gibbs potential is given by (follows from Eq. (\ref{reacq}))

\begin{equation}
  \label{gibbs}
g(\sigma)=\frac{\sigma^2}{2\rho^0 E}-\frac{\sigma}{\rho}.
\end{equation}
In Fig. \ref{figphase} we show an equilibrium phase diagram in the
conjugate pair of variables $\sigma$ and $1/\rho$. If the derivative
of the Gibbs potential with respect to the external field $\sigma$
is evaluated at the critical point $\sigma=0$, it can be seen that
there are two possible scenarios. The first scenario is a first
order phase transition, which occurs whenever there is a jump in the
referential densities, i.e. the derivative of the Gibbs potential is
discontinuous and the second derivative diverges at the critical
point. The other scenario is a second order phase transition, which
occurs when the referential densities of the two phases are
identical. We then have a jump in the second order derivative
whenever Young's modules of the two phases are dissimilar.

The order of the phase transition has a fundamental impact on the
dynamics. In two dimensions a first order phase transition kinetics
will generally lead to morphological instabilities of the
propagating phase boundary while a second order phase transition
will either flatten or roughen the boundary depending on Poisson's
ratios of the two materials. In the next section we analyze the
different phase transitions by performing a linear stability
analysis.

\section{Linear perturbation analysis}
We now solve the elasto-static Eqs. (\ref{vdot}) and
(\ref{forcejump}) together with the kinetics Eqs. (\ref{nvec}) and
(\ref{massex}) in two dimensions for an arbitrary perturbation to an
initially flat interface using the quasi-static version of momentum
balance in Eq. (\ref{vdot}). In addition to the translational
dynamics observed in the one-dimensional system presented above, it
turns out, that in two dimensions the interface dynamics is
non-trivial and may lead to the formation of finger-like structures.
The general setup is shown in Fig. \ref{2dsetup} where phase $i$,
$i=1,2$, has material parameters $\mu_i, \nu_i$ and $\rho_i$, with
$\mu_i$ being the shear modulus and $\nu_i$ being the Poisson's
ratio. In general, the interface velocity depends on its morphology,
the 6 material parameters and the external loading
$\sigma_{\infty}$. One degree of freedom is removed by rescaling the
shear modulus of one phase with the external load.

\subsection{Stress field around a perturbed flat interface}
In order to evaluate the jump in Gibbs energy density, i.e.
$\bl\mathcal{F}/\rho^0+\mathcal{W}\br$, we need to determine the stress
field around the interface by solving the elastostatic equations. We
have that under plane stress conditions, the local strain energy
density can be written on the form
\begin{equation}\label{S3-Energy}
\mathcal{F} =\frac{1}{4\mu}\biggl(\sigma_{xx}^2+
\sigma_{yy}^2-\frac{\nu}{1+\nu}(\sigma_{xx}+\sigma_{yy})^2+2\sigma_{xy}^2\biggr)
\end{equation}
and the work term is defined as
\begin{equation}\label{S3-work}
\mathcal{W} =-\sigma_{nn}\rho_i^{-1}= -\sigma_{nn}\rho_{i,0}^{-1}(1+\mathrm{Tr}(\mathbf\epsilon)).
\end{equation}
The trace of strain is given in terms of stress by
\begin{equation}
\mathrm{Tr}({\bf \epsilon}) = \frac{1-2\nu}{2\mu(1+\nu)}(\sigma_{xx}+\sigma_{yy}).
\end{equation}
Note that we could as well have formulated the problem under plane
strain conditions; however, the generic behavior in both plane
stress and strain is the same although the detailed dependence on
the material parameters is altered.

We solve the mechanical problem by finding the Airy stress
function, $U(x,y)$ \cite{Musk53}, which satisfies the biharmonic
equation $\Delta^2 U = 0$. Once the stress function has been found,
the stress tensor components readily follow from the relations
\begin{gather}\label{S3-1}
\sigma_{xx} = \frac{\partial^2 U}{\partial y^2}, \quad \sigma_{yy} =
\frac{\partial^2 U}{\partial x^2}, \quad \sigma_{xy} =
-\frac{\partial²U}{\partial x\partial y}.
\end{gather}
The biharmonic equation is solved under the boundary conditions of a
normal load applied in the y direction at infinity, i.e.
$\sigma_{yy}\rightarrow-|\sigma_{\infty}|<0$ and $\sigma_{xy}=0$ for
$y\rightarrow\pm\infty$. The continuity of the
stress vector across the interface follows from force balance. In
addition we require that $u_x(\pm \infty,y)=0$.

For a flat interface, the stress field is homogenous in space. This
implies that the Airy stress function is quadratic in $x$ and $y$, with coefficients determined by the
boundary conditions. With the boundary conditions specified above,
the stress function for the i-th phase can be written in the form
\begin{equation}
U_i(x,y) = \frac{|\sigma_{\infty}|}{2}(x^2+\nu_i y^2).
\end{equation}

From this stress function we can calculate the Gibbs potential which in
the case of dissimilar phases is discontinuous across the
interface. The velocity of the phase transformation readily follows
from the potential 
\begin{eqnarray}
W_0 &\propto& \bl\mathcal{F}_0/\rho^0+\mathcal{W}_0\br
=|\sigma_{\infty}|\left(\frac{1}{\rho^0_1}-
\frac{1}{\rho^0_2}\right)\nonumber\\
&-&\frac{|\sigma_{\infty}|^2}{4}\left(\frac{1-3\nu_1}{\rho^0_1\mu_1}-
\frac{1-3\nu_2}{\rho^0_2\mu_2}\right).\label{unperturbed}
\end{eqnarray}
The subscript of the free energy density and the work term refers to an
unperturbed interface. From the above equation, we see that when the lower phase is much
denser than the upper phase, i.e.~$\rho^0_1\ll\rho^0_2$, the
interface propagates uniformly into the upper phase with a velocity
$W\approx|\sigma_{\infty}|\bl 1/\rho\br>0$, i.e. the denser phase
grows into the softer. When the densities are identical or almost
identical, $\rho_2/\rho_1\approx 1$ and the shear modules
significantly different, i.e. $\mu_1\ll\mu_2$. When the two solids
phases have identical Poisson's ratios, $\nu$, we see that the softer phase can only grow into the harder one when $\nu<1/3$.

In the case of an arbitrarily shaped interface separating the two
phases, the analytical solution to the stress field is in general
far from trivial. In-plane problems can in some cases be solves using
conformal mappings or perturbation schemes~\cite{Musk53,Gao91,Regev08}. Here, we solve the stress field
around a small undulation of flat interface employing a linear perturbation
scheme~\cite{Gao91}. In the linear stability analysis we now study the
growth of an arbitrary harmonic perturbation with wavelength $k$, i.e. $h(x,t) =
Ae^{\omega t}\cos(kx)$ with $A\ll 1$. In appendix \ref{genpert}, we
derive expressions for a general perturbation. The Airy
stress function can be written as a
superposition of the solution to the flat interface and a small
correction due to undulation, $U(x,y) = U_0(x,y)+\Theta(x,y)$, where $\Theta(x,y)$ is determined from the
interfacial constraints of continuous stress vector and displacement
field. When the wave number $k$ is much smaller than the cutoff introduced by the surface tension, we obtain the
following expressions for the Airy stress functions
\begin{eqnarray}
\Theta_1(x,y) &=& \frac{-|\sigma_\infty|h(x)\exp(-ky)(\alpha_1 y+\beta)}{k(\mu_2\kappa_1+\mu_1)(\mu_1\kappa_2+\mu_2)}\nonumber\\
\Theta_2(x,y) &=& \frac{|\sigma_\infty|h(x)\exp(ky)(\alpha_2 y-\beta)}{k(\mu_2\kappa_1+\mu_1)(\mu_1\kappa_2+\mu_2)}\label{airy}
\end{eqnarray}
where $\kappa_i = \frac{3-\nu_i}{1+\nu_i}$ and we have introduced the material specific constants,
\begin{eqnarray*}
\alpha_1&=&k(1-\nu_1)(\mu_2-\mu_1)(\mu_1\kappa_2+\mu_2)\\
\alpha_2&=&k(1-\nu_2)(\mu_1-\mu_2)(\mu_2\kappa_1+\mu_1)
\end{eqnarray*}
and
\begin{eqnarray*}
\beta = 2\mu_1^2\frac{1-\nu_2}{1+\nu_2}-2\mu_2^2\frac{1-\nu_1}{1+\nu_1}+4\mu_1\mu_2\frac{\nu_1-\nu_2}{(1+\nu_2)(1+\nu_1)}\\
\end{eqnarray*}

From the Airy stress functions, we then calculate the stress components using Eq.
(\ref{S3-1}) and find the jumps in the Gibbs energy density from Eqs.
(\ref{S3-Energy}) and (\ref{S3-work}). The evolution of the shape
perturbation relative to a uniform translation of the flat interface
is described by Eq. (\ref{massex}), namely
\begin{equation}
\frac{\partial h(x,t)}{\partial t}\propto \bl\mathcal{F}+\mathcal{W}\br-W_0,
\end{equation}
which in the linear regime corresponds to a dispersion relation
given in the general form as
\begin{equation}\label{dispersion}
\omega \propto \frac{\bl\mathcal{F}+\mathcal{W}\br-W_0}{h}.
\end{equation}
Below follows an evaluation of the growth rate for a small harmonic
perturbation to a flat interface. For this perturbation, the general
expression for the growth rate follows directly upon insertion of the
Airy functions in Eq.~(\ref{airy}) and then in Eq.~(\ref{S3-1}), however, the growth rate is not easily expressed in a short and readable form and we have therefore limited our presentation to a few special cases.
The growth rate is a function of the six material parameters ($\nu_i,\mu_i,\rho_i$) and the external stress. Naturally, the stability of the growing interface is invariant under the interchange of the solid phases and correspondingly the region of the stability diagram that we have to study is reduced. 
\subsection{First and second order phase transition: Stability diagrams}
\begin{figure}[t]
\epsfig{file=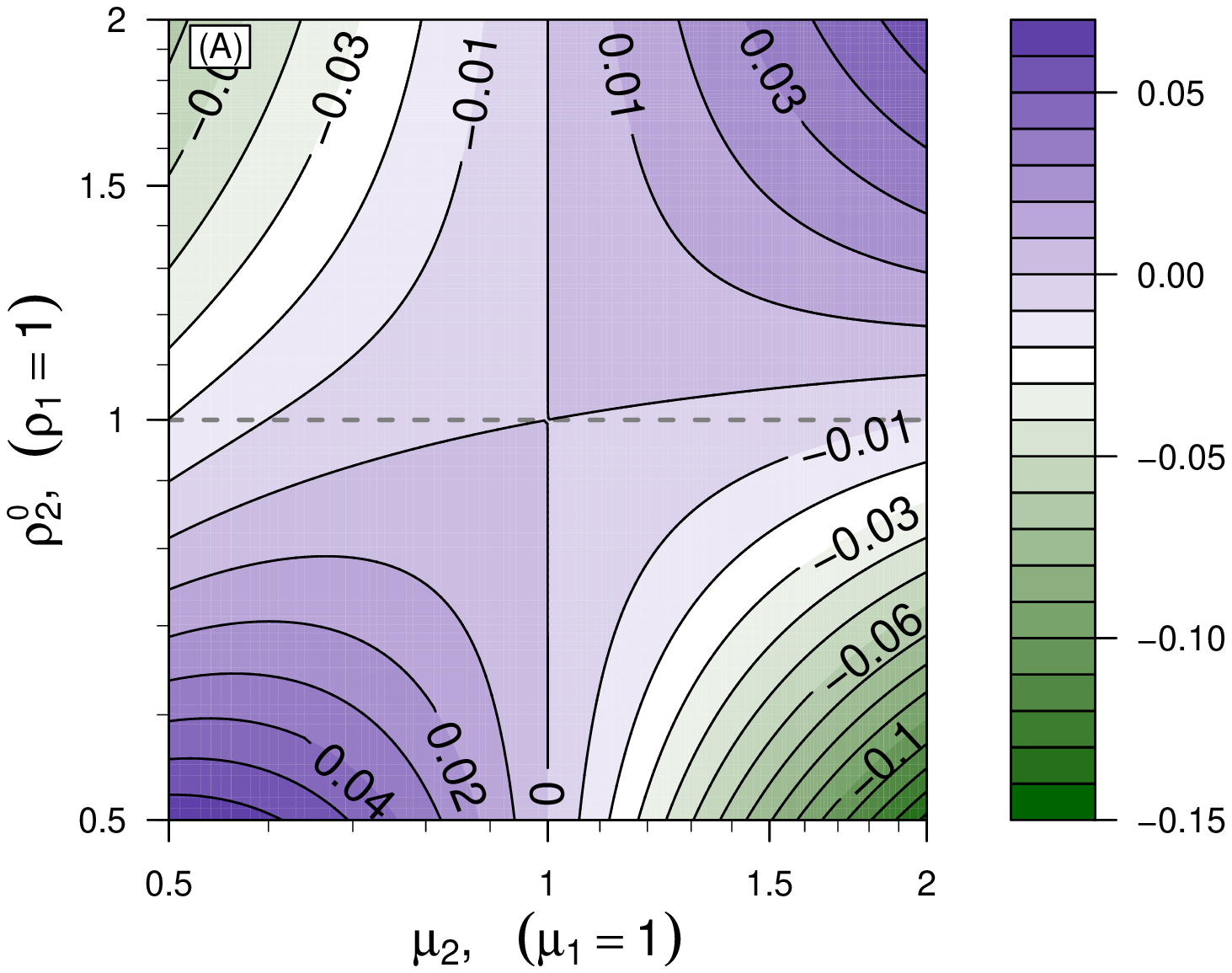,width=.47\textwidth}
\epsfig{file=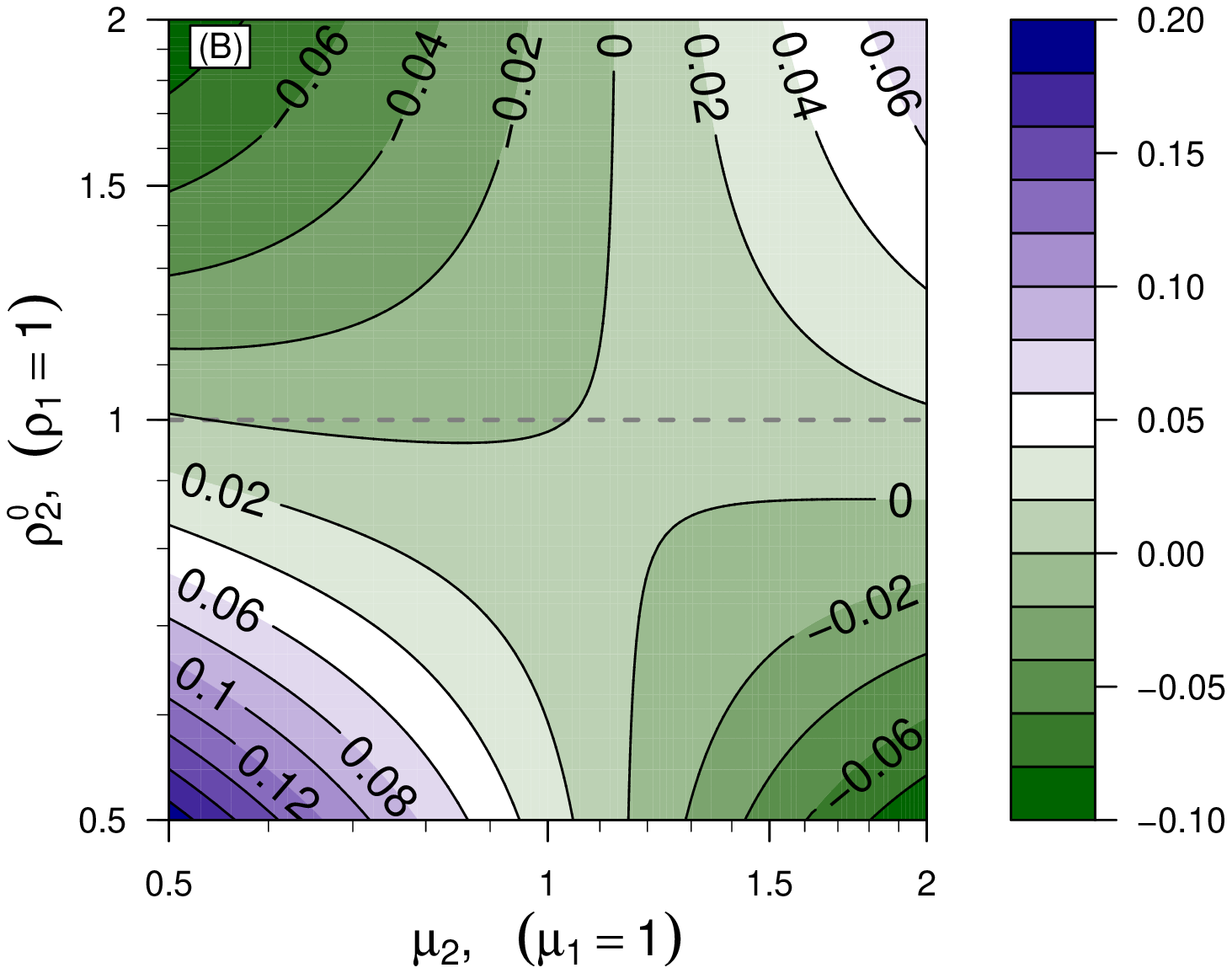,width=.47\textwidth}
  \caption{(color online) Panel (A), stability diagram for two solids materials with
identical Poisson's ratio of $\nu=0.25$. Panel (B), diagram for
solids with Poisson's ratios of $\nu_1=0.45$ and
$\nu_2=0.40$.}\label{Fig:contour-diag}
 \end{figure}

In the second order phase transition when both solids have the same
referential densities $\rho_1^0=\rho_2^0=\rho^0$ and when the
Poisson's ratios $\nu_1=\nu_2=\nu$ are identical the dispersion relation assumes a simple form
\begin{equation}
\frac \omega k =\frac{(3\nu-1)(1-\nu)(\mu_1+\mu_2)(\mu_2-\mu_1)^2}{\rho^0\mu_1\mu_2(\mu_1+\mu_2\kappa)(\mu_2+\mu_1\kappa)(1+\nu)}\label{2nd}
\end{equation}
where $\kappa$ is the fraction introduced above and $k$ the wave number of the perturbation.  The expression reveals an interesting behavior where the interface is stable for Poisson's ratio less than 1/3 and is unstable for Poisson's ratio larger than 1/3. Fig. (\ref{Fig:contour-diag}) shows stability diagrams for the specific case where $\mu_1=1$ and $\rho_1^0=1$ (in arbitrary units). In panel (A) the diagram is calculated for two solids that have the same Poisson's ratio and with a value $\nu=1/4$. The second order phase transition occurs along the horizontal cut $\rho_2^0=1$ and is marked by a dashed grey line. We observe that $\omega/k$ is negative along this line and the interface is therefore stable. For $\nu$ larger than $1/3$ (not shown in the figure) the horizontal zero level curve will flip around and the grey dashed line will then be covered with unstable regions. In order to see this flip, we expand Eq. (\ref{dispersion}) around the point (1,1), i.e.~in terms of $\rho_2^0-1$ and $\mu_2-1$, and achieve the following expression for the zero curve 
\begin{equation}\label{2ndstab}
\rho_2^0\approx 1+\frac{(1-2\nu-3\nu^2)(\mu_2-1)}{\nu (7+\nu)}
\end{equation}
Note that the right hand side is in units of $\rho_1$. We directly observe that the horizontal zero curve flips around at the critical point $\nu=1/3$.
In the case when the two solids are identical, i.e. at the point (1,1) in the stability diagram, all modes will as expected remain unchanged and the interface therefore remain unaltered. The other parts of the zero levels lead to marginal stability but will in general induce a growth of the interface due to the unperturbed Gibbs potential Eq.~(\ref{unperturbed}).
We now consider a cut in the stability diagram where the two solids have the same shear modules, $\mu_1=\mu_2=\mu$, but different densities and Poisson's ratios. For different Poisson's ratios the dispersion relation Eq. (\ref{dispersion}) becomes 
\begin{equation}
\frac \omega k = \frac{(\nu_2-\nu_1)(\nu_1\rho^0_2-\nu_2\rho_1^0+2(\rho_2^0-\rho_1^0)\mu)}{4\rho_1^0\rho^0_2\mu}
\end{equation}
From this expression we see that the vertical zero line observed in Eq.~(\ref{2nd}) and in Fig. \ref{Fig:contour-diag} panel (A) only exists for identical Poisson's ratios. When the solids have different Poisson's ratios, the separatrix or intersection of the two zero curves located at (1,1) in panel (A) will split into two non-intersecting zero curves. In panel (B) we show a stability diagram for solids with Poisson's ratios $\nu_1=0.45$ and $\nu_2=0.40$.

In general the stability diagram is characterized by four quadrants,
two stable and two unstable, delimited by neutral zero curves. The
physical regions would typically correspond to the quadrants $I$ and
$III$ under the assumption that higher density implies higher shear
modulus. In these quadrants the growth rate is typically positive
(i.e. the interface is unstable) except for a thin region at the
borderline between a first and second order phase transition, i.e. when $\rho_2\simeq \rho_1$.

\section{Numerical results and discussions}

The linear stability analysis revealed an intricate change in
stability depending on the material properties and densities of the
two solids. We explore this stability beyond the linear regime using
numerical methods. The bulk elasto-static equation Eq.(\ref{vdot})
is solved numerically on an unstructured triangular grid using the
Galerkin finite element method and the surface tension force is
converted to a body force in a narrow band surrounding the
interface. The discontinuous jumps appearing in the dynamical Eq.
(\ref{massex}) are computed at the outer border of the band.
Periodic boundary conditions are used to minimize the possible
influence of the finite system size in the x-direction (parallel to
the interface). The interface is tracked using a level set method
(e.g.~\cite{Sethian99}) and propagated with the normal velocity
calculated in Sec. II using Eq. (\ref{massex}). Several level set
functions, $\phi(x,t)$, can be used, however, most level set methods
use the signed distance function ($|\phi(x,t)|$ is the shortest
distance between $x$ and the interface and the sign of $\phi(x,t)$
identifies the phase at position $x$). Good numerical accuracy can
be obtained by keeping $\phi(x,t)$ a signed distance function at all
times, and this is achieved by frequent reinitialization of
$\phi(x,t)$ according to the iterative scheme
\begin{equation}\label{eq:ls4}
\frac{\partial\phi}{\partial t'} + S(\phi_0)(|\nabla\phi|-1) = 0,
\end{equation}
where $\phi_0$ is the level set function before the reinitialization, $t'$ is a
fictitious time,  and $S( \phi_0 ) = \phi_0/\sqrt{\phi_0^2 + (\Delta
x)^2}$, where $\Delta x$ is the grid size. Generally only a couple
of iterations are needed at each time step, to obtain a good
approximation to a signed distance function, and it is only necessary
to update the level set function in a narrow band around the interface.

\begin{figure}
 \epsfig{width=.45\textwidth,file=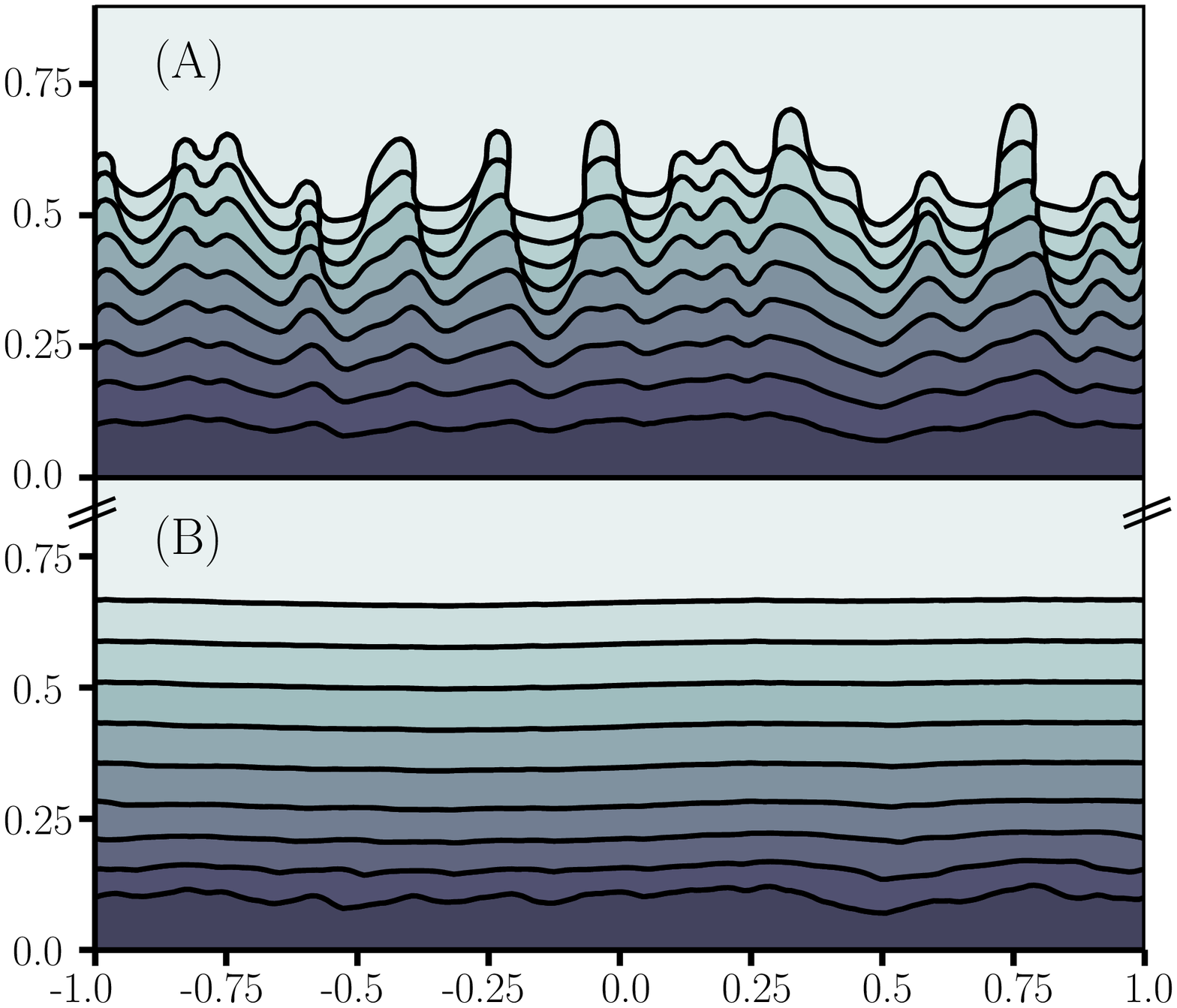}
 \epsfig{width=.48\textwidth,file=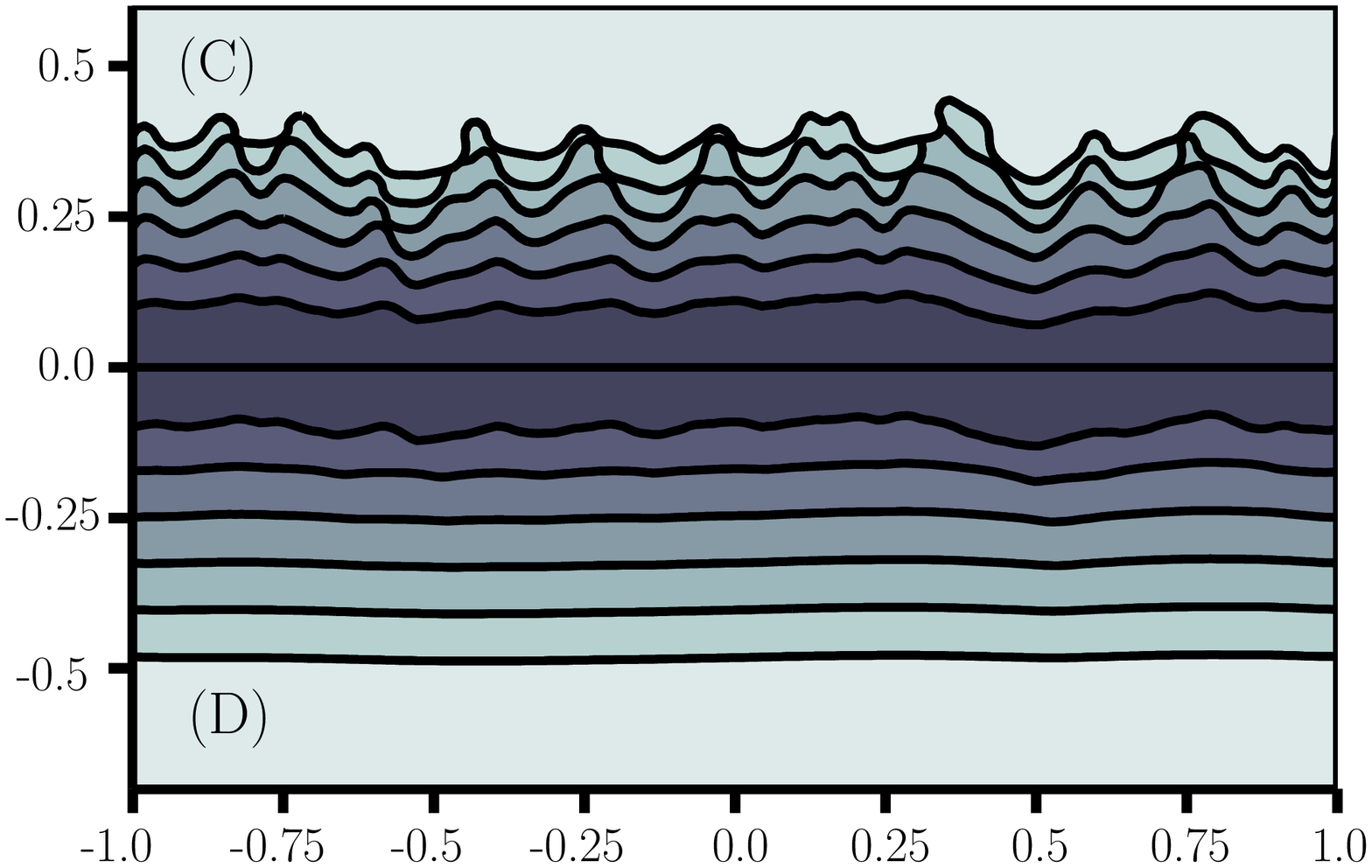}
 \caption{\label{fig:profile}(color online) Simulations of the
 temporal evolution of solid-solid interfaces for first order (Panels
 (A)  and (B)) and second order (Panels (C) and (D)) phase transitions. Panel (A) shows a simulation using
$\rho_1=1.0, \mu_1=1.0$ and $\rho_2=1.05,\mu_2=2.0$. Both phases
have identical Poisson's ratio, $\nu_1=\nu_2=0.45$. Panel (B) is a
simulation run with densities and shear modules similar to panel (A)
but with a different Poisson's ratios, $\nu_1=\nu_2=0.25$. Panel (C) is a simulation run with  $\rho_1=1.0,
\mu_1=1.0$ and $\rho_2=1.0,\mu_2=2.0$. Both phases have identical
Poisson's ratios, $\nu_1=\nu_2=0.45$. Panel (D) shows a simulation run
 with densities and shear modules similar to Panel (C) but with
different Poisson's ratios, $\nu_1=\nu_2=0.25$. The color code
 represents a time arrow pointing from the darker blue (early stage)
 to the lighter blue (final stage).}
 \end{figure}

\begin{figure}
 \epsfig{width=.45\textwidth,file=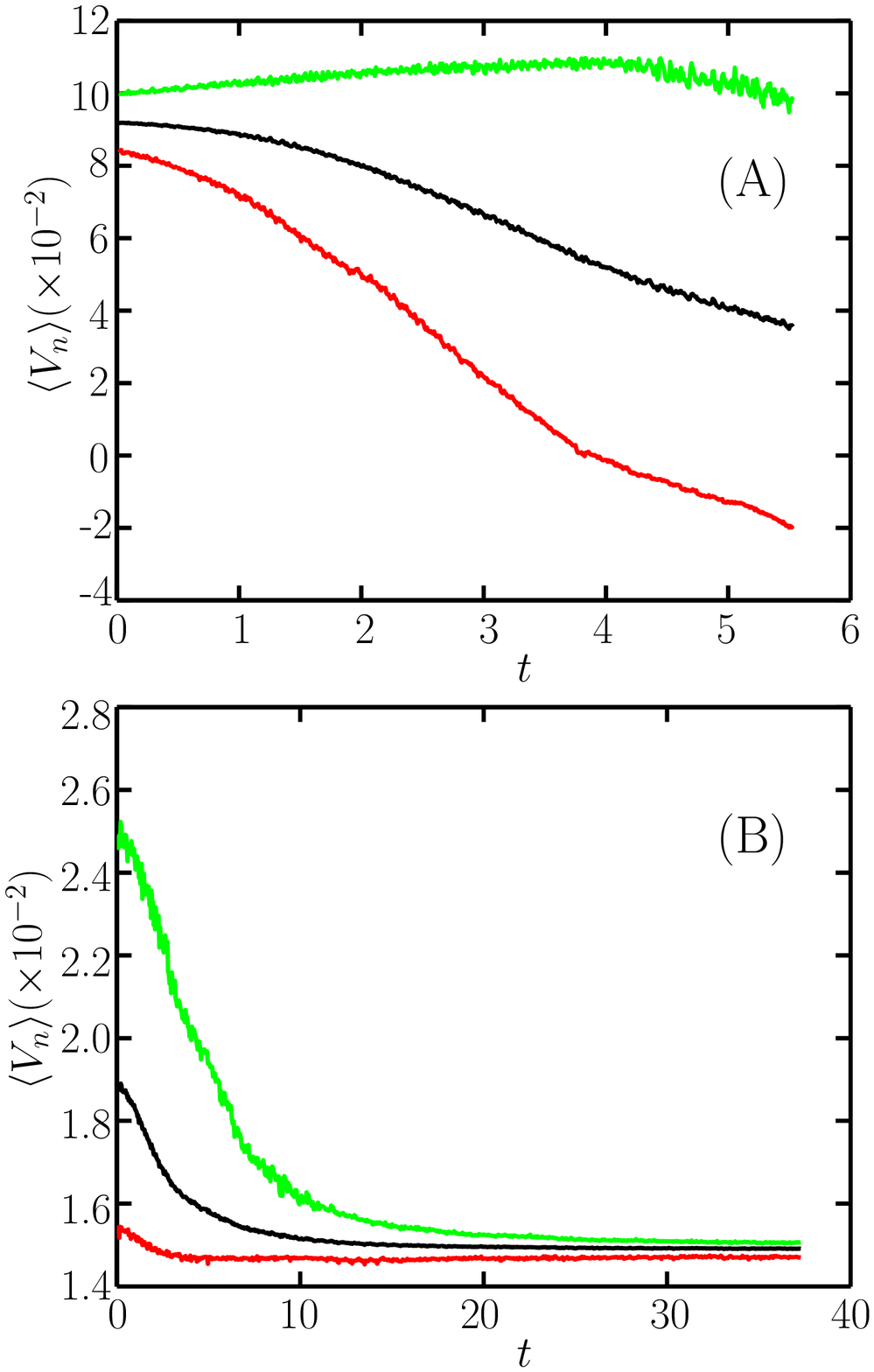}
 \epsfig{width=.45\textwidth,file=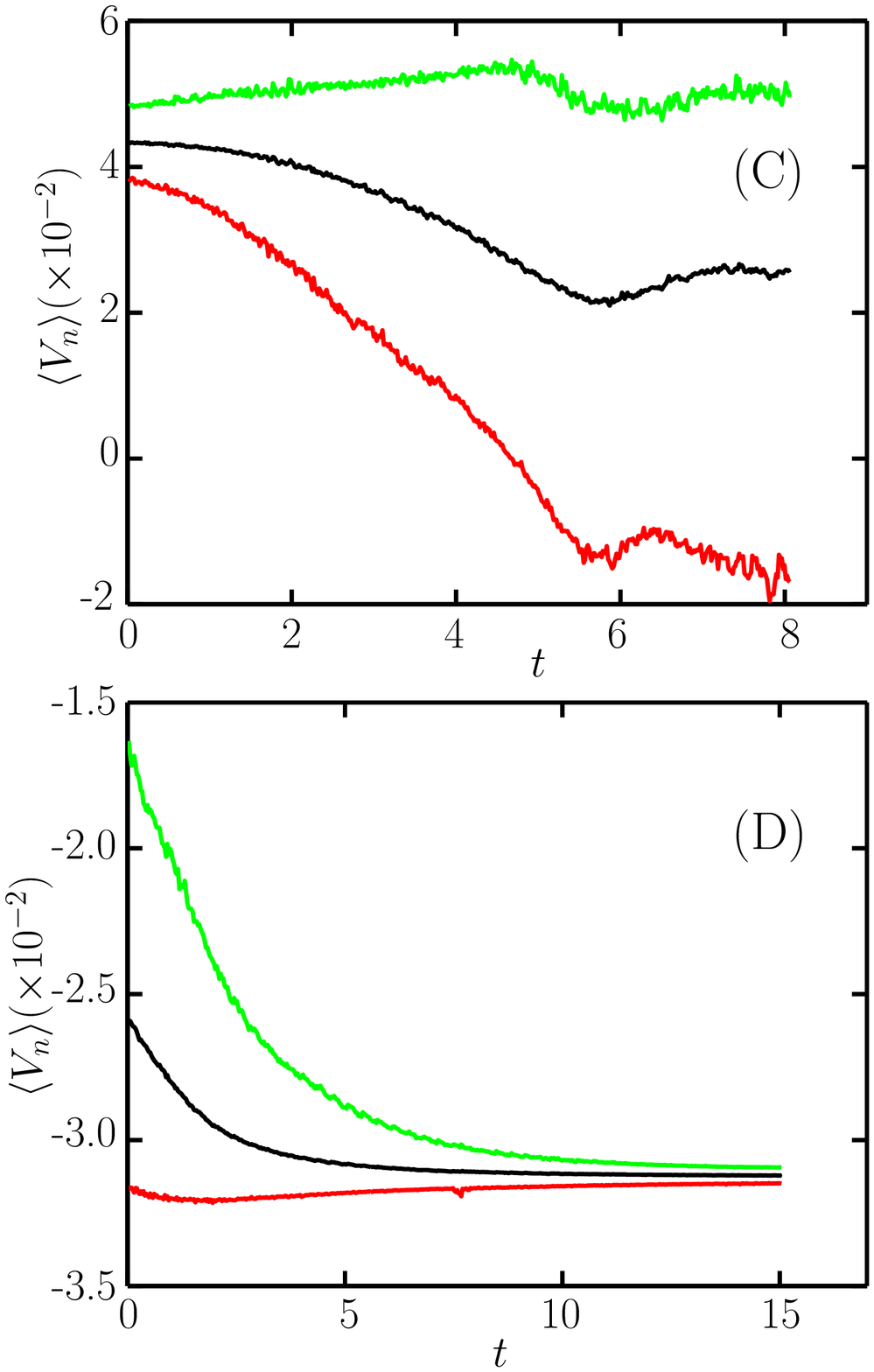}
 \caption{\label{fig:mean-vel}(color online) Normal velocity as a
 function of time in first order (Panels (A) and (B)) and second order
 (Panels (C) and (D)) phase transitions. The simulations presented in the
 individual panels are identical to the corresponding panels in
 Fig. \ref{fig:profile}. The color code of the curves reference to the
 mean velocity (in black), the mean of lowest $10$ \% (in red) and the mean of the highest $10$ \% (in green).}
 \end{figure}

In Figs.~(\ref{fig:profile}) and (\ref{fig:mean-vel}) we present
numerical simulations of the phase transformation kinetics using
parameter regions where the interface is either stable or unstable. The simulations presented in
Fig.~(\ref{fig:profile}) panels (A) and (B) represent interface snap
shots of a
first order phase transition dynamics and panels (C) and (D) simulations of a second
order, respectively. In panel (A), the values
of the parameters were chosen in a region of the stability diagram where the
interface is predicted to roughen and in panel (B) we have used
parameters corresponding to a stable
evolution of the interface. Note that the interface in both cases is moving from the dense
phase into the soft phase independent of its stability. This is in
agreement with the one dimensional calculation performed in
Sec. II. Panel (C) shows a
case of a second order phase transition where the interface is
unstable, while panel (D) shows a stable case. We notice that, for second order phase
transitions, the overall translation of the interface is
changed in unison with its stability. In Eq.~(\ref{2ndstab}) we saw that the
stability of the second order phase transition is dictated by
the values of Poisson's ratios. For Poisson's ratio smaller
than $1/3$, the kinetics is stable and the phase of small shear modulus grows into the phase
of higher shear modulus while for higher values of Poisson's ratio
the behavior is reversed and the interface roughens with time. This
also follows from Eq. (\ref{unperturbed}). In fig.~\ref{fig:mean-vel}, we have plotted the mean velocity as a
function of time for the simulations presented in fig.~\ref{fig:profile}.

\section{Concluding remarks}
In conclusion, it has been shown that the phase transformation of one
solid into the another across a thin interface may lead to a
morphological instability, as well as the development of fingers along
the propagating interface. We have presented a stability analysis
based on the Gibbs potential for non-hydrostatically stressed solids
and have established a linear relationship between the rate of entropy
production at the interface and the rate of mass exchange between the
solid phases. The solids are compressed transverse to the interface
and corresponding stability diagrams reveal an intricate dependence of
the stability on the material density, Poisson's ratio and Young's
modulus. With the density as order parameter, two types of phase
transitions were considered, a first and a second order, respectively. 

For both types of transitions we find expressions for the curves
separating the stable and unstable regions in the stability
diagram. For most material parameters the first order phase
transition, i.e. when the two solids have different referential
densities, destabilizes the interface by allowing fingers to grow from
the denser phase into the other. When the solids have identical or
almost identical densities, i.e. a second order phase transition, we
find that the stability depends on Poisson's ratios of the two solids.
If the two solids have Poisson's ratios less than 1/3, the phase
transition dynamics of the two solids will lead to a flattening of the
interface, i.e. any perturbation of a flat interface will decay and
ultimately the interface will propagate uniformly from the soft phase
(low Young's modulus) into the hard phase (high Young's modulus).

We believe that our classification of the phase transition order
together with the stability analysis may find application in many
natural systems, since the morphological stability directly provide
information about the order of the underlying phase transformation
process and the material parameters.

\begin{acknowledgments}
This project was funded by \textsl{Physics of Geological Processes},
a Center of Excellence at the University of Oslo. The authors are
grateful to R. Fletcher, P. Meakin, Y.Y. Podladtchikov, and F.
Renard for fruitful discussions and comments.
\end{acknowledgments}
\appendix
\section{Surface tension}
In this appendix we present additional details on the derivation of
the reaction rate Eq. (\ref{massex}) including the interfacial free
energy.  Let us consider a diffuse interface characterized by a
small thickness over which the concentration field varies smoothly
between the constant values in the bulk of the two phases. In the
Cahn-Hilliard formalism, the free energy is introduced as a function
of both the concentration and concentration gradients, and has the
form
\begin{equation}\label{barf}
\rho\bar f(\bar\epsilon_{ij},c,\nabla c) = \rho\bar f_0(\bar\epsilon_{ij},c)+\frac{\kappa_1}{2}|\nabla c|^2,
\end{equation}
where the first term is the free energy in the bulk and the second
term is associated with the interfacial free energy. Here $\kappa_1$
is a small parameter related to the thickness of the interface.

In this case, the calculation of the reaction rate $Q$ proceeds as
in Sec. II. We apply the total time derivative of the local
equilibrium equation, Eq. (\ref{bare}), where the free energy is
given by Eq. (\ref{barf}) and then obtain the following expression
\begin{equation}
\dot{\bar e} = \frac{\partial\bar f}{\partial\bar\epsilon_{ij}}\dot{\overline{\epsilon_{ij}}}+\frac{\partial\bar f}{\partial c}\dot c+\frac{\partial \bar f}{\partial\nabla_i c}\left(\nabla_i \dot c-\nabla_j c\nabla_i \bar v_j\right)+T\dot{\bar s},
\end{equation}
where the commutation relation, $\frac{d}{dt}\nabla_i c = \nabla_i \dot c-\nabla_i v_j\nabla_j c$, has been used \cite{Trusk98}. Combining the above equation with the conservation of energy from Eq. (\ref{edot}) and the entropy balance from Eq.(\ref{sdot}) an expression for the entropy production rate is obtained
\begin{eqnarray*}
T\Pi_s &=& \left(\sigma_{ij}+\rho\nabla_jc\frac{\partial\bar f}{\partial\nabla_i c}\right)\nabla_i \bar v_j -\rho\left(\frac{\partial\bar f}{\partial c}-\nabla_i\frac{\partial\bar f}{\partial\nabla_i c}\right)\dot c-\rho\frac{\partial\bar f}{\partial\bar\epsilon_{ij}}\dot{\overline{\epsilon_{ij}}}\\
&=& n_i\left(\sigma_{ij}+\rho \nabla_ic\frac{\partial\bar f}{\partial\nabla_j c}\right)n_jQ\delta_\Gamma\frac{\partial}{\partial c}\left(\frac{1}{\tilde\rho}\right)-\left(\frac{\partial\bar f}{\partial c}-\nabla_i\frac{\partial\bar f}{\partial\nabla_i c}\right)Q\delta_\Gamma\\
&+&\left(\sigma_{ij}+\rho\nabla_ic\frac{\partial\bar f}{\partial\nabla_j c}-\rho\frac{\partial\bar f}{\partial\bar\epsilon_{ij}}\right)\dot{\overline{\epsilon_{ij}}}.
\end{eqnarray*}
We observe that $\Pi_s$ satisfies the second law of thermodynamics
provided that the last term vanishes and the rest of the terms are
brought into a quadratic form. This implies a constitutive equation
for the stress given by
\begin{equation}\label{stress}
\sigma_{ij} = \rho\frac{\partial\bar f}{\partial\bar\epsilon_{ij}}-\rho\nabla_ic\frac{\partial\bar f}{\partial\nabla_j c},
\end{equation}
and a linear kinetics law with the reaction rate being proportional to
\begin{equation}
Q\approx K\left(\rho\frac{\partial\bar f_0}{\partial\bar\epsilon_{ij}}n_in_j\frac{\partial}{\partial c}\left(\frac{1}{\rho}\right)-\frac{\partial \bar f}{\partial c}+\nabla_i\frac{\partial\bar f}{\partial \nabla_i c}\right),
\end{equation}
where $K$ is a positive local constant of proportionality and
$\sigma_{ij}^0$ is the elastic stress in the absence of surface
tension.

Using Eq.~(\ref{barf}), the two constitutive laws may be expressed as
\begin{eqnarray}
\sigma_{ij} & =& \sigma^0_{ij}-\kappa_1\nabla_ic\otimes \nabla_j c\\
Q&=& K\left(\sigma^0_{nn}\frac{\partial}{\partial c}\left(\frac{1}{\rho}\right)-\frac{\partial \bar f_0}{\partial c}+\kappa_1\rho^{-1}\nabla^2c \right),
\end{eqnarray}
where $\sigma_{ij}^0$ is the elastic stress obtained in Sec. II without the surface stress.

In the sharp interface limit, i.e. the thickness goes to zero, the surface free energy becomes
\begin{equation}
\rho f^{surf} = \kappa_1|\nabla c|^2 \rightarrow \gamma \delta_\Gamma,
\end{equation}
and surface stress is related to the surface energy by
\begin{equation}
\sigma_{ij}^{surf} = \kappa_1|\nabla c|^2\left(1-\frac{\nabla_i\phi}{|\nabla \phi|}\otimes\frac{\nabla_j\phi}{|\nabla \phi|}\right)\rightarrow \gamma (1-n_i\otimes n_j)\delta_\Gamma.
\end{equation}
The divergence of the surface stress is then calculated as
\begin{equation}
\nabla_i\sigma^{surf}_{ij}=2\mathcal{K}\gamma n_j\delta_\Gamma,
\end{equation}
where $\mathcal{K}$ is the local curvature.

\section{Goursat functions around a perturbed flat interface}\label{genpert}
In this appendix, we explain in details how to calculate the Airy stress
functions around the perturbed flat interface introduced in Sec.
III. All the detailed calculations were carried out in Maple in
order to handle the lengthy algebraic expressions.

The Airy stress function satisfies the biharmonic equation $\partial^2_z\partial^2_{\overline z} U = 0$. This equation has a general solution which can be written in the Goursat form $U(z,\overline z) = \Re\{\overline
z\phi(z)+\chi(z)\}$, where $\varphi(z)$ and $\chi(z)$ are complex functions determined by the boundary conditions. Combining Eq. (\ref{S3-1}) with the Goursat solution, stress components are related to these functions by the following expressions
\begin{eqnarray}
\sigma(z) &=& \sigma_{xx}(x,y)+\sigma_{yy}(x,y) = 2\{\varphi'(z)+\overline{\psi'}(z)\},\label{S3-2} \\
\Sigma(z) &=& \sigma_{yy}(x,y)-\sigma_{xx}(x,y)+2i\sigma_{xy}(x,y) \nonumber\\
&=&2\{\overline{z}\varphi''(z)+\psi(z)\}\label{S3-3},
\end{eqnarray}
where $\varphi(z) = \chi'(z)$.
The solution to the biharmonic equation is determined up to a linear gauge transformation,
\begin{eqnarray}
&&\varphi(z)\mapsto \varphi(z)+Ciz+p\\
&&\psi(z)\mapsto\psi(z)+q,
\end{eqnarray}
where $C$ is a real number and $p$, $q$ are arbitrary complex numbers.

The boundary conditions are given by the far-field stresses and the
constraints at the interface. Here we consider that the system is
loaded by a uniaxial compression in the y-direction,
$\sigma_{yy}(x,\infty) =-|\sigma_{\infty}|<0$. Whenever the two
phases are different an interface is introduced at which we require
force balance and continuous displacement field. The force balance
is expressed by the following jump condition
\begin{equation*}
\bl \sigma_{xx}n_x+\sigma_{xy}n_y+i(\sigma_{yx}n_x+\sigma_{yy}n_y)\br=-
\gamma\mathcal{K}(n_x+in_y),
\end{equation*}
where $\mathcal{K}$ is the local curvature and $\gamma$ is the
surface tension. From Eqs. (\ref{S3-2}) and (\ref{S3-3}) we find
that the force balance leads to the following condition on the
Goursat functions
\begin{equation}\label{S3-stressjump}
\bl\varphi+z\overline{\varphi'}+\overline{\psi}\br=
i\int_{0}^{s}\gamma\mathcal{K}(n_x+in_y)ds,
\end{equation}
where $s$ is a point at the interface. The continuity of the
displacement field across the interface introduces an additional
jump condition given by
\begin{equation}\label{S3-displjump}
\bl\frac{1}{\mu}(-\kappa\varphi+z\overline{\varphi'}+\overline{\psi})\br =0,
\end{equation}
where $\mu$ is the shear modulus and $\kappa = \frac{3-\nu}{1+\nu}$
is a constant for in-plane stress-elasticity determined by the
Poisson's ratio.

The two jump conditions, Eqs. (\ref{S3-stressjump}) and
(\ref{S3-displjump}) combined with the far-field boundary
conditions, $\varphi_\infty(z) =
-\frac{1}{4}(1+\nu)|\sigma_\infty|z$ and
$\psi_\infty(z)=-\frac{1}{2}(1-\nu)|\sigma_\infty|z$ are sufficient
to determine the fields $\varphi_1(z)$, $\psi_1(z)$, $\varphi_2(z)$
and $\psi_2(z)$.

Superimposing an arbitrary perturbation with amplitude $h(x)$ on the flat interface, the Goursat functions are slightly altered. They can be expanded to linear order in $h(x)$ as follows \cite{Gao91},
\begin{eqnarray}
\varphi(x)&\approx& \varphi_0(x)+i h(x)\varphi_0'(x)+\Phi(x)\\
\psi(x)&\approx& \psi_0(x)+i h(x)\psi_0'(x)+\Psi(x).\\
\end{eqnarray}
$\Phi(x)$ and $\Psi(x)$ are functions of $h(x)$. Inserting this expansion into Eqs. (\ref{S3-displjump}) and (\ref{S3-stressjump}), we obtain that the corresponding jump conditions for the perturbation fields
\begin{eqnarray}
\bl\Phi(x)+x\overline{\Phi'}(x)+\overline{\Psi}(x)\br &=&
ih(x)\bl\overline{\Sigma_0}(x)\br\nonumber\\
&+&f(x) \label{S3-PerturbJump1}\\
\bl \frac{-\kappa\Phi(x)+x\overline{\Phi'}(x)+\overline{\Psi}(x)}{\mu}\br
&=& ih(x)\bl\frac{\overline{\Sigma_0}(x)}{\mu}\br\label{S3-PerturbJump2},
\end{eqnarray}
where $f(x) = i\int_0^x\gamma\mathcal{K}(n_x+in_y)ds$. To linear order we find that $f(x)\approx-\gamma\int_0^xh''(s)ds$. Eqs. (\ref{S3-PerturbJump1}) and (\ref{S3-PerturbJump2}) can be rewritten equivalently as
\begin{eqnarray}
&&\Phi_1(x)-\Omega\biggl(x\overline{\Phi'_1(x)}+\overline{\Psi_1(x)}\biggr)-(1+\Lambda)\Phi_2(x)\nonumber\\
&&=-i\Omega h(x)\overline{\Sigma}_{01}(x)+\frac{1+\Lambda}{1+\kappa}f(x)\label{App-Jump1}\\
&&\Phi_2(x)-\Pi\biggl(x\overline{\Phi'_2(x)}+\overline{\Psi_2(x)}\biggr)-(1+\Delta)\Phi_1(x)\nonumber\\
&&=-i\Pi h(x)\overline{\Sigma}_{02}(x)-\frac{1+\Delta}{1+\kappa}f(x).\label{App-Jump2}
\end{eqnarray}
The constants appearing above are expressed in terms of the elastic moduli. Adopting the notation of \cite{Gao91}, these are given by
\begin{gather}
\Lambda = \kappa\frac{1/\mu_2-1/\mu_1}{1/\mu_2+\kappa/\mu_1}, \qquad \Pi = \frac{1/\mu_2-1/\mu_1}{\kappa/\mu_2+1/\mu_1}\\
\Delta = \kappa\frac{1/\mu_1-1/\mu_2}{\kappa/\mu_2+1/\mu_1},\qquad
\Omega = \frac{1/\mu_1-1/\mu_2}{\kappa/\mu_1+1/\mu_2}.
\end{gather}
Eqs. (\ref{App-Jump1}) and (\ref{App-Jump2}) are solved at an arbitrary point $z$ in the complex plane by applying the Cauchy integral and using the analytic continuation of each function \cite{Musk53}. Let us denote the Cauchy integral over the perturbation amplitude
\begin{eqnarray}
H_1(z) &=& \frac{1}{2\pi i}\int\frac{h(x)}{x-z}dx,\textrm{ with }\Im(z)>0\\
H_2(z) &=& \frac{1}{2\pi i}\int\frac{h(x)}{x-z}dx,\textrm{ with }\Im(z)<0.
\end{eqnarray}
Notice that the two functions satisfy the following relations
\begin{gather*}
\overline{H_1(\overline{z})} = -H_2(z),\qquad \overline{H_2(\overline{z})} = -H _1(z)\\
\Im(H_1(x))=\Im(H_2(x)),\qquad \Re(H_1(x))=-\Re(H_2(x)),
\end{gather*}
where the principal value of the Cauchy integral is considered when $x$ is a point on the real axis.

Thus, by applying the Cauchy integral with $\Im(z)>0$ in Eq. \ref{App-Jump1} and $\Im(z)<0$ in Eq. \ref{App-Jump2}, $\Phi_1$ and $\Psi_2$ are determined in the integral form as follows
\begin{eqnarray*}
\Phi_1(z) &=& -i\Omega\Sigma_{0,1}H_1(z)+\frac{1+\Lambda}{1+\kappa}F_1(z)\\
\Phi_2(z) &=& i\Pi\Sigma_{0,2}H_2(z)+\frac{1+\Delta}{1+\kappa}F_2(z),\\
\end{eqnarray*}
where
\begin{equation}
F'(z)=\frac{1}{2\pi i}\int\frac{f'(x)}{x-z}dx \approx
-\gamma\frac{d^2}{dz^2}H(z).
\end{equation}
$\Psi_1(z)$ is calculated from the complex conjugation of Eq. (\ref{App-Jump1}) when the Cauchy integral is applied on both sides of the equation and $\Im(z)>0$. In a similar manner, $\Phi_2(z)$ is derived from Eq. (\ref{App-Jump2}). The final expressions for the two functions then follow
\begin{eqnarray*}
\Psi_1(z) &=&-i\Sigma_{0,1}H_1(z)-\frac{1+\Lambda}{1+\kappa}\biggl(-i\Pi\Sigma_{0,2}H_1(z)\\
&&-\frac{1+\Delta}{1+\kappa}F_1(z)\biggr)-\frac{1+\Lambda}{\Omega(1+\kappa)}F_1(z)-z\Phi'_1(z)\\
\Psi_2(z) &=&
i\Sigma_{0,2}H_2(z)-\frac{1+\Delta}{\Pi}\biggl(i\Omega\Sigma_{0,1}H_2(z)\\
&&-\frac{1+\Lambda}{1+\kappa}F_2(z)\biggr)-\frac{1+\Delta}{\Pi(1+\kappa)}F_2(z)-z\Phi'_2(z).
\end{eqnarray*}

For a cosine perturbation of the interface, $h(x) = A \cos(kx)$,
with $A\ll 1$ the Airy stress function, $U(x,y) = \Re\{\bar
z\phi(z)+\chi(z)\}$ is obtained explicitly. 
%

%
%

\end{document}